\documentclass[12pt]{article}
\usepackage{geometry}                % See geometry.pdf to learn the layout options. There are lots.
\usepackage{amsmath}
\usepackage{graphicx}
\usepackage{amssymb}
\usepackage{epstopdf}
\newcommand{\bds}{\boldsymbol}

\newcommand{\PRD}{{\it Phys. Rev. D\ }}
\newcommand{\PRL}{{\it Phys. Rev. Lett.\ }}
\newcommand{\PR}{{\it Phys. Rev. \ }}
\newcommand{\NUCL}{{\it Nucl. Phys.\ }}

\title{Persistence of zero modes in a gauged\\
Dirac model for bilayer graphene}
\author{\small R. Jackiw\\[-.5ex]
\it \small Department of Physics\\[-.5ex]
\it \small Massachusetts Institute of Technology\\[-.5ex]
\it \small Cambridge, MA 02139\\[2ex]
\small S.-Y. Pi\\[-.5ex]
\it \small Department of Physics\\[-.5ex]
\it \small Boston University\\[-.5ex]
\it \small Boston, MA 02215\\
 \tt  \small CTP-MIT/3970
}
\date{}                                           % Activate to display a given date or no date

\begin{document}
\maketitle
\begin{abstract}
A recently constructed model for low lying excitations in bilayer graphene exhibits mid-gap, zero energy modes in its Dirac-like spectrum, when a scalar order parameter takes a vortex profile. We show that these modes persist when the dynamics is extended by a gauge field interaction, which also renders finite the vortex energy. The effect of the gauge field on the zero energy wave function is to shift the phase of the (damped) oscillatory component of the wave function in the absence of the gauge field.
\end{abstract}
%\section{}
%\subsection{}

%\section*{Persistence of zero modes in a gauged\\
%Dirac model for bilayer graphene}

\indent The old subject of Dirac zero modes and fractional charge \cite{jack072008-1} revived recently owing to the emergence of graphene  as an experimentally realized planar substance \cite{jack072008-2}, whose low-energy excitations can be described by a Dirac equation in two spatial dimensions \cite{jack072008-3}. If the material exhibits various dimerization patterns, the effective Dirac fields also interact with a homogenous scalar field (order parameter), and this gives rise to a gap in the Dirac spectrum. When the scalar field acquires a topologically interesting profile, {\it e.g.} a vortex, a zero energy, mid-gap state can occur with fractional (fermion) charge \cite{jack072008-1}.

An early instance of planar Dirac zero modes was found in \cite{jack072008-4}, but no actual experimental setting was given. Today graphene, and graphene-like substances, offer the possibility of a physical realization.

Monolayer graphene consists of a hexagonal, honeycomb lattice, which may be presented as a superposition of two triangular sublattices, A and B. In the tight-binding approximation, there are two Dirac points. If a particular dimenization ---called Kekul\'{e} distortion --- occurs,  the effective Dirac Hamiltonian also possesses an interaction with a scalar field $\varphi$ \cite{jack072008-5}. 
\begin{eqnarray}
h_1 &=& \bds{\alpha \cdot p} + \beta\, |\varphi| \ e^{-i \gamma_{5}\chi}\label{rjackiw:eq1} \\[1ex]
\text{Here} && \bds{\alpha} = \begin{pmatrix}
     \bds{\sigma} & 0    \\
      0&\bds{-\sigma}  
\end{pmatrix},\quad
\beta = \begin{pmatrix}
     0 & I    \\
     I &  0
\end{pmatrix}, \quad
\gamma_{5} = \begin{pmatrix}
   I   &  0  \\
    0  &  -I \nonumber
\end{pmatrix}\\ [1ex]
&& \bds{p} = \frac{1}{i}\, \bds{\nabla}, \quad \varphi = |\varphi| e^{i \chi} \nonumber
\end{eqnarray}
The $4 \times 4$  Dirac Hamiltonian $h_1$ acts on a $4 $-spinor $\Psi$,
\begin{equation}
\Psi = \left(\begin{array}{c}\psi^B_+ \\[1ex] \psi^A_+ \\[1ex] \psi^A_- \\[1ex] \psi^B_-\end{array}\right)
\label{rjackiw:eq2}
\end{equation}
where ($\pm$) refer to the two Dirac points and ($A, B$) label the sublattices. The vectorial quantities $\bds{p}, \bds{\alpha} \ \text{and}\ \bds{\sigma}$ are 2-dimensional. The kinetic term $\bds \alpha \cdot \bds p$ does not mix the Dirac points; mixing arises through $\varphi$ as a consequence of the Kekul\'{e} distortion. Homogenous $\varphi = m$ produces  a mass gap, while an $n$-vortex profile for $\varphi ({\bf r}) $ produces zero modes.  The Hamiltonian \eqref{rjackiw:eq1} anti-commutes with $\alpha^{3} = \left(\begin{array}{cc}\sigma^3 & 0 \\0 & -\sigma^{3}\end{array}\right)$.
\begin{equation}
\alpha^{3} h_1 \alpha^{3} = -h_1
\label{rjackiw:eq3}
\end{equation}
Therefore $\alpha^{3}$ maps positive energy solutions  onto negative energy solutions and zero modes can be chosen to be eigenmodes of $\alpha^{3}$. This ``energy reflection symmetry" is a manifestation of the sublattice symmetry found in the honeycomb graphene lattice.

The above model was extended by including an interaction with a gauge field, $\bf A$, whose purpose is to unpin the vortices; in the absence of the gauge field the scalar vortex carries infinite energy (per unit legnth) \cite{jack072008-6}.
\begin{equation}
h^{\bf A}_{1} = \bds{ \alpha} \cdot (\bds{p} - \gamma_{5}\,  \mathbf{ A}) + \beta\, |\varphi | \, e^{-i \gamma_{5}\chi} 
\label{rjackiw:eq4}
\end{equation}
The gauged model, possesses a local chiral gauge symmetry
\begin{alignat}{3}
& \Psi \to e^{i \omega \gamma_5 } \Psi, \quad \varphi \to e^{2 i \omega} \varphi \Rightarrow \chi \to \chi + 2 \omega \nonumber\\
&\mathbf{ A} \to \mathbf{ A} + \bds{\nabla} \omega ,
\label{rjackiw:eq5}
\end{alignat}
and one readily verifies the identity
\begin{eqnarray}
h^{\mathbf{ A}}_{1} &=& \text{exp}\, \left(\alpha^3\, \frac{1}{\nabla^{2}}\ b \right) \ h_1\  \text{exp} \ \left(\alpha^3\, \frac{1}{\nabla^{2}}\ b \right),\label{rjackiw:eq6}\\[1ex]
b &=& \varepsilon^{ij}\, \partial_i\  A^{j}.
\label{rjackiw:eq7}
\end{eqnarray}
Thus the extended model still retains the energy reflection symmetry, and possess zero energy eigenmodes, whose wave functions differ from those with just a scalar vortex by the factor $e^{- \alpha^{3} \, \frac{1}{\nabla^2}\, b}$.

However, it may be difficult to achieve experimentally the Kekul\'{e} distortion. Recently a model that is physically different but mathematically similar to \eqref{rjackiw:eq1} has been put forward, with the suggestion that the excitation condensate, needed for topological effects, fractional charge etc., can ``be produced in the laboratory in the near future" \cite{jack072008-7}. The physical system consist of a graphene bilayer, separated by a dielectic barrier, and biased by an external, constant voltage $V$. In a mean-field approximation, the Hamiltonian for the above bilayer system is given by
\begin{equation}
h_{2} = \bds{\alpha} \cdot \bds{p} + \beta\, | \varphi | \, e^{- i \gamma_5 \chi} + \gamma_5 \, V
\label{rjackiw:eq8} 
\end{equation}
 which acts on the $4$-spinor $\Psi$. %$h_{2}$
\begin{equation}
\Psi = \left(\begin{array}{c}\psi^{B}_{1} \\[1ex] -\psi^{A}_{1} \\[1ex] \psi^{B}_{2} \\[1ex] \psi^{A}_{2} \end{array}\right)
\label{rjackiw:eq9}
\end{equation}
As before ($A, B$) refer to the sublattices, but (\ref{rjackiw:eq1}, \ref{rjackiw:eq2}) label the two layers, which are nested, one directly above the other.  There are no Dirac point labels, because the above description refers to a single Dirac point in each lattice of the two stacked lattices. Here $\varphi$ describes the condensate arising from states bound by interlayer Coulomb forces between particles in one layer and holes in the other. This dynamics is modeled by a 4-Fermi interaction of strenght $U$. A gap equation is solved in the Hartee-Fock approximation, leading to
\begin{equation}
|\varphi | \approx \sqrt{\wedge V}\ e^{-\sqrt{3}\, \pi\, t^{2}/U V}  .
\label{rjackiw:eq10}
\end{equation}
Here $t$ is the hopping amplitude between sites on each of the two monolayers; there is no interlayer hopping within this model's approximations. Eq. \eqref{rjackiw:eq10} holds in the limit $\wedge \gg V \gg | \varphi |$, where $\wedge$ is an energy cut off \cite{jack072008-7}. It is striking that this order parameter enters the bilayer theory in a way identical to the Kekul\'{e} distortion of the monolayer model, $h_1$ in \eqref{rjackiw:eq1}.
 
 The presence of $\gamma_{5} V$ in $h_{2}$, which has no analog in $h_1$, spoils the energy reflection symmetry \eqref{rjackiw:eq3}. But another property of $h_{2}$ ensures similar behavior. One verifies that $h_{2}$ satisfies 
 \begin{equation}
\beta \, \alpha^{2}\,  h^{\ast}_{2} \ \beta\, \alpha^{2}  = h_2.%\psi^{\ast}_{E}
\label{rjackiw:eq11}
\end{equation}
Thus energy reflection works as
\begin{equation}
\Psi_{- E} = \beta\, \alpha^2\ \Psi^\ast_E ,
\label{rjackiw:eq12}
\end{equation}
and $h_{2}$ possesses zero-energy eigenstates, satisfying \cite{jack072008-7}
\begin{equation}
\alpha^2\, \beta\, \Psi_0 = \Psi^\ast_0 .
\label{rjackiw:eq13}
\end{equation}

In view of our earlier work on gauging the monolayer graphene model \cite{jack072008-6}, we are led to study the gauged version of $h_{2}$. 
\begin{equation}
h^{\mathbf{ A}}_{2} = \bds{\alpha} \cdot (\bds{p} - \gamma_5\, \mathbf{ A}) + \beta\, | \varphi | \, e^{-i \gamma_5 \, \chi} + \gamma_5\, V
\label{rjackiw:eq14}
\end{equation}
Gauge transformations follow \eqref{rjackiw:eq5} and $V$ is gauge invariant. The new energy reflection property, \eqref{rjackiw:eq11}, \eqref{rjackiw:eq12}, is maintained. Consequently we expect to find zero modes, which we now exhibit .

The 4-spinor \eqref{rjackiw:eq9} is presented in terms of 2-spinors.
\begin{equation}
\Psi= \left(\begin{array}{c}\Psi_1 \\[1ex] \Psi_2\end{array}\right),  \Psi_1 = \left(\begin{array}{c}\psi^B_1 \\[1ex] -\psi^A_1\end{array}\right) , \Psi_2 = \left(\begin{array}{c}\psi^B_2 \\[1ex] \psi^A_2\end{array}\right)
\label{rjackiw:eq15}
\end{equation}
With our Dirac matrices, the zero energy spinors satisfy according to \eqref{rjackiw:eq13}, $\Psi_2 = \sigma^2\, \Psi^\ast_1$ and the eigenvalue equation reads
\begin{subequations}\label{rjackiw:eq16ab}
\begin{eqnarray}
(\bds\sigma \cdot (\bds{p}- \mathbf{ A}) + V )\ \Psi_1 + \varphi\, \sigma^2 \, \Psi^\ast_1 &=& 0 \label{rjackiw:eq16a}\\[1ex]
\varphi^\ast\, \Psi_1 - (\bds\sigma \cdot (\bds{p}- \mathbf{ A}) + V ) \ \sigma^2 \, \Psi^\ast_1 &=& 0
\label{rjackiw:eq16b}
\end{eqnarray}
\end{subequations}

In fact, the second equation is a consequence of the first, and needs not be considered separately.  Continuing with the matrix reduction, we set $\Psi_1 (\mathbf{ r}) = \left(\begin{array}{c}F(\mathbf{ r}) \\ G(\mathbf{ r})\end{array}\right) $, and eq. \eqref{rjackiw:eq16a} now reads %\varphi (\mathbf{ r}) = m (\mathbf{ r}) \, e^{i n \theta}
\begin{subequations}
\begin{eqnarray}
V G - i \, e^{i \theta} \left[\partial_+ + \frac{n}{r}\ A \right] \ F + i\, m\, e^{i n \theta} \, F^\ast &=& 0 \label{rjackiw:eq17a},\\[1ex]
V F - i \, e^{-i \theta} \left[\partial_- - \frac{n}{r}\ A \right]\ G - i\, m\, e^{i n \theta} \, G^\ast &=& 0\label{rjackiw:eq17b},\\[1ex]
\partial_\pm \equiv \frac{\partial}{\partial r} \pm i\, \frac{i}{r}\ \frac{\partial}{\partial\theta}, \hspace{.75in}\nonumber
\end{eqnarray}
\end{subequations}
where we have taken $A^i  = - n \, \varepsilon^{ij}\ \frac{r^j}{r^{2}}\ A\, (r), \text{with}\ A (0) = 0, A(\infty) = \frac{1}{2}$ and $\varphi = m\, (r) \, e^{i n \theta}, \text{with}\ m\, (0) = 0, m (\infty) = m$. 

To separate the angular dependence, to make the equations real and to simplify them, we posit the {\it Ansatz}
\begin{subequations}
\begin{eqnarray}
F (\mathbf{ r}) &=& -i \frac{f(r)}{\sqrt{r}}\ e^{-(M (r) - i l_1 \theta)},\label{rjackiw:18a}\\[1ex]
G(\mathbf{ r}) &=& \frac{g(r)}{\sqrt{r}}\ e^{-(M (r) - i l_2 \theta)},
\label{rjackiw:18b}
\end{eqnarray}
\end{subequations}
$l_1 = \frac{n-1}{2}, l_2 = \frac{n+1}{2},  M^\prime (r) = m (r) \ \text{and}\ f, g$ are real. Single valuedness requires that $n$ be an odd integer. The final equations read
\begin{subequations}\label{rjackiw:eq19}
\begin{eqnarray}
&\Rightarrow&\left(\partial_r - \frac{n}{r}\ \left(\frac{1}{2} - A\right)\right)\ f - V g = 0,\label{rjackiw:eq19a}\\[1ex]
&\Rightarrow&\left(\partial_r +\frac{n}{r}\ \left(\frac{1}{2} - A\right)\right)\ g + V f = 0.
\label{rjackiw:eq19b}
\end{eqnarray}
\end{subequations}

When $A$ remains unspecified (apart from its asymplotes) eqs. \eqref{rjackiw:eq19}  do not appear explicitly integrable. Nor can $A$ be removed, as in the monolayer case \eqref{rjackiw:eq6}, \eqref{rjackiw:eq7}. However, one can show that a normalizable solution exists.

For $r \to \infty, A \to \frac{1}{2}$ and \eqref{rjackiw:eq19} reduce to
\begin{subequations}\label{rjackiw:eq20}
\begin{alignat}{2}
&f^\prime \ -V g =0,\label{rjackiw:eq20a}\\[1ex]
&g^\prime\ + V f = 0,\label{rjackiw:eq20b}
\end{alignat}
\end{subequations}
with solution that involves two constants, ($c, d$).
\begin{subequations}\label{rjackiw:eq21}
\begin{alignat}{2}
&f (r) = c \cos V r + d\, \sin V r \label{rjackiw:eq21a}\\[1ex]
&g (r) = -c\, \sin  V r + d\, \cos V r \label{rjackiw:eq21b}
\end{alignat}
\end{subequations}
Evidently owing to the $r^{-\frac{1}{2}} \ e^{- M (r)}$ factor both $F$ and $G$ are always damped at large $r$. Thus the wave function will be acceptable and normalizable if a solution that is regular at the origin can be constructed.

At the origin $A$ vanishes and the equations \eqref{rjackiw:eq19} reduce to
\begin{subequations}\label{rjackiw:eq22}
\begin{alignat}{2}
&\left( \frac{\partial}{\partial r} - \frac{n}{2 r}\right)\ f - V g = 0\label{rjackiw:eq22a}\\[1ex]
&\left(\frac{\partial}{\partial r} + \frac{n}{2 r}\right)\ g + V f = 0\label{rjackiw:eq22b}
\end{alignat}
\end{subequations}
Of course these are the same equations, which hold for all $r$ in the absence of $\mathbf{ A}$, as with the Hamiltonian $h_{2}$ in \eqref{rjackiw:eq8}. Their solution is given in terms of Bessel functions \cite{jack072008-7}.
\begin{equation}
f = r^{\frac{1}{2}}\, J_{\frac{n}{2} - \frac{1}{2}}\ (V r), \ g = -r^{\frac{1}{2}}\ J_{\frac{n}{2} + \frac{1}{2}}\ (V r)
\label{rjackiw:eq23}
\end{equation}
Note that the large r asymptote of \eqref{rjackiw:eq23} is of the form \eqref{rjackiw:eq21} with specific values for $c = \sqrt{\frac{2}{\pi}}\ \cos\ \frac{n \pi}{4}, d = \sqrt{\frac{2}{\pi}}\ \sin\ \frac{n \pi}{4}$.  Thus the effect of the gauge field is to move $c$ and $d$ from the above values; {\it i.e.} $\bf A$ causes a phase shift in the profiles without gauge field. 

We acknowledge important conversations with A. Castro-Neto,  C. Chamon, G. Semenoff and B. Seradjeh. This work was supported by the Department of Energy under contract No. DE-FG02-05ER41360 and No. DE-FG02-91ER40676. The research was performed at the Aspen Center for Physics.


\begin{thebibliography}{9}
\bibitem{jack072008-1}
R. Jackiw and C. Rebbi, \PRD {\bf 13}, 3398 (1976); W.-P. Su, J.R. Schrieffer and J. Heeger, \PRL {\bf 42}, 1698 (1979); R. Jackiw and J.R. Schrieffer, \NUCL {\bf B190}, 253 (1981).

\bibitem{jack072008-2}
K.S. Novoselov {\it et. al.}, {\it Science} {\bf 306}, 666 (2004).

\bibitem{jack072008-3}
P. R. Wallace, \PR {\bf 71}, 622 (1947); G. Semenoff, \PRL {\bf 53}, 2449 (1984).

\bibitem{jack072008-4}
R. Jackiw and P. Rossi, \NUCL {\bf 190}, 681 (1981).

\bibitem{jack072008-5}
C.-Y. Hou, C. Chamon and C. Mudry, \PRL {\bf 98}, 186809 (2007).

\bibitem{jack072008-6}
R. Jackiw and S.-Y. Pi, \PRL {\bf 98}, 266402 (2007).

\bibitem{jack072008-7}
B. Seradjeh, H. Weber and M. Franz, arXiv 0806.0849 (cond-mat).


\end{thebibliography}
\end{document}